\documentclass[10pt,conference]{IEEEtran}
\usepackage[T1]{fontenc}
\usepackage{hyperref}
\hypersetup{pdftex,colorlinks=true,allcolors=blue}
\usepackage{tcolorbox}
\newtcolorbox{takeaway}{colframe=black,colback=gray!15,boxrule=1pt,arc=2pt,left=2pt,right=2pt,top=1pt,bottom=1pt, before skip=1em, after skip=0em}
\usepackage{url}
\usepackage[normalem]{ulem}

\AtBeginDocument{%
	\providecommand\BibTeX{{%
			\normalfont B\kern-0.5em{\scshape i\kern-0.25em b}\kern-0.8em\TeX}}}

\usepackage{setspace}
\usepackage{longtable}
\usepackage[figuresright]{rotating}
\usepackage{epsfig,endnotes,color}
\usepackage{graphicx}
\usepackage{amsmath}
\usepackage{caption}
\usepackage{graphicx}
\usepackage{url}
\usepackage{xcolor}
\usepackage{transparent}
\usepackage{float}
\usepackage{lipsum}
\usepackage{listings}
\definecolor{codegreen}{rgb}{0,0.6,0}
\definecolor{codegray}{rgb}{0.5,0.5,0.5}
\definecolor{codepurple}{rgb}{0.58,0,0.82}
\definecolor{backcolour}{rgb}{0.95,0.95,0.95}
\definecolor{verylightgray}{rgb}{.97,.97,.97}
\definecolor{custompink}{RGB}{216,100,139}

\lstdefinelanguage{Solidity}{
	keywords=[1]{anonymous, assembly, assert, balance, break, call, callcode, case, catch, class, constant, continue, constructor, contract, debugger, default, delegatecall, delete, do, else, emit, event, experimental, export, external, false, finally, for, function, gas, if, implements, import, in, indexed, instanceof, interface, internal, is, length, library, log0, log1, log2, log3, log4, memory, modifier, new, payable, pragma, private, protected, public, pure, push, require, return, returns, revert, selfdestruct, send, solidity, storage, struct, suicide, super, switch, then, this, throw, transfer, true, try, typeof, using, value, view, while, with, addmod, ecrecover, keccak256, mulmod, ripemd160, sha256, sha3}, 
	keywordstyle=[1]\color{blue}\bfseries,
	keywords=[2]{address, bool, byte, bytes, bytes1, bytes2, bytes3, bytes4, bytes5, bytes6, bytes7, bytes8, bytes9, bytes10, bytes11, bytes12, bytes13, bytes14, bytes15, bytes16, bytes17, bytes18, bytes19, bytes20, bytes21, bytes22, bytes23, bytes24, bytes25, bytes26, bytes27, bytes28, bytes29, bytes30, bytes31, bytes32, enum, int, int8, int16, int24, int32, int40, int48, int56, int64, int72, int80, int88, int96, int104, int112, int120, int128, int136, int144, int152, int160, int168, int176, int184, int192, int200, int208, int216, int224, int232, int240, int248, int256, mapping, string, uint, uint8, uint16, uint24, uint32, uint40, uint48, uint56, uint64, uint72, uint80, uint88, uint96, uint104, uint112, uint120, uint128, uint136, uint144, uint152, uint160, uint168, uint176, uint184, uint192, uint200, uint208, uint216, uint224, uint232, uint240, uint248, uint256, var, void, ether, finney, szabo, wei, days, hours, minutes, seconds, weeks, years},	
	keywordstyle=[2]\color{teal}\bfseries,
	keywords=[3]{block, blockhash, coinbase, difficulty, gaslimit, number, timestamp, msg, data, gas, sender, sig, value, now, tx, gasprice, origin},	
	keywordstyle=[3]\color{violet}\bfseries,
	identifierstyle=\color{black},
	sensitive=false,
	comment=[l]{//},
	morecomment=[s]{/*}{*/},
	commentstyle=\color{codegreen}\ttfamily,
	stringstyle=\color{red}\ttfamily,
	morestring=[b]',
	morestring=[b]"
}

\lstdefinelanguage{SolidityNoKeywords}{
  keywords=[1]{},
  keywords=[2]{},
  keywords=[3]{},
  sensitive=false
}

\newcommand{\solstatement}[1]{%
  \lstinline[
    language=SolidityNoKeywords,
    basicstyle=\color{custompink},
    keywordstyle=\color{custompink},
    stringstyle=\color{custompink},
    commentstyle=\color{custompink},
    identifierstyle=\color{custompink},
    numberstyle=\color{custompink},
  ]|#1|%
}

\usepackage{svg}
\usepackage{CJKutf8}
\usepackage{subfigure}
\usepackage{lscape, latexsym, amssymb, algorithmic, multirow}
\usepackage[linesnumbered, vlined, ruled]{algorithm2e}
\usepackage{multirow}
\usepackage{mathtools, bbm, color}
\usepackage{booktabs}
\usepackage{amsthm,mathrsfs,amsfonts,dsfont}
\usepackage{enumerate}
\usepackage{hhline}

\usepackage{enumitem}
\usepackage{tabu}
\usepackage{colortbl}
\usepackage[graphicx]{realboxes}

\usepackage{footnote}
\usepackage{color}
\usepackage{colortbl}
\usepackage{pifont}
\usepackage{bbding}
\usepackage{threeparttable}
\usepackage{rotating}
\usepackage{adjustbox}

\usepackage{marvosym}

\usepackage{minted}
\setminted{
  breaklines=true, 
  breakanywhere=true 
}

\def\BibTeX{{\rm B\kern-.05em{\sc i\kern-.025em b}\kern-.08em
    T\kern-.1667em\lower.7ex\hbox{E}\kern-.125emX}}

\newcommand{\system}{{\sc PromFuzz}\xspace}

\IEEEoverridecommandlockouts

\newfont{\mycrnotice}{ptmr8t at 7pt}

\title{\system: Leveraging LLM-Driven and Bug-Oriented Composite Analysis for Detecting Functional Bugs in Smart Contracts}

\author{
\IEEEauthorblockN { 
Xingshuang Lin\IEEEauthorrefmark{1}\thanks{${\textrm{\ding{70}}}$: Xingshuang Lin and Qinge Xie are the co-first authors.}$^{\textrm{\ding{70}}}$,
Qinge Xie\IEEEauthorrefmark{2}$^{\textrm{\ding{70}}}$,
Binbin Zhao\thanks{${\textrm{\Letter}}$: Binbin Zhao is the corresponding author.}\IEEEauthorrefmark{1}$^{(\textrm{\Letter})}$\IEEEauthorrefmark{6},
Yuan Tian\IEEEauthorrefmark{3},
Saman Zonouz\IEEEauthorrefmark{2},
Na Ruan\IEEEauthorrefmark{4},
Jiliang Li\IEEEauthorrefmark{5}, \\
Raheem Beyah\IEEEauthorrefmark{2}, 
Shouling Ji\IEEEauthorrefmark{1}
}
\IEEEauthorblockA {
  \IEEEauthorrefmark{1}Zhejiang University,
  \IEEEauthorrefmark{2}Georgia Institute of Technology, 
  \IEEEauthorrefmark{3}University of California, Los Angelos,
  \IEEEauthorrefmark{4}Shanghai Jiaotong \\ University, 
  \IEEEauthorrefmark{5}Xi'an Jiaotong University,
  \IEEEauthorrefmark{6}Engineering Research Center of Blockchain Application, Supervision And \\ Management (Southeast University), Ministry of Education
}

\IEEEauthorblockA {
  E-mails: 
  cs.xslin@zju.edu.cn, 
  qxie47@gatech.edu, 
  binbinz@zju.edu.cn, 
  yuant@ucla.edu, 
  szonouz6@gatech.edu, \\
  naruan@cs.sjtu.edu.cn,  
  jiliang.li@xjtu.edu.cn, 
  rbeyah@coe.gatech.edu, 
  sji@zju.edu.cn
}
}

\begin{document}

\IEEEaftertitletext{\vspace{-1.0\baselineskip}}
\maketitle
\begin{abstract} 
Smart contracts are fundamental pillars of the blockchain, playing a crucial role in facilitating various business transactions. However, these smart contracts are vulnerable to exploitable bugs that can lead to substantial monetary losses. A recent study reveals that over $80\%$ of these exploitable bugs, which are primarily functional bugs, can evade the detection of current tools. Automatically identifying functional bugs in smart contracts presents challenges from multiple perspectives. The primary issue is the significant gap between understanding the high-level logic of the business model and checking the low-level implementations in smart contracts. Furthermore, identifying deeply rooted functional bugs in smart contracts requires the automated generation of effective detection oracles based on various bug features.

To address these challenges, we design and implement \system, an automated and scalable system to detect functional bugs in smart contracts. In \system, we first propose a novel Large Language Model (LLM)-driven analysis framework, which leverages a dual-agent prompt engineering strategy to pinpoint potentially vulnerable functions for further scrutiny. We then implement a dual-stage coupling approach, which focuses on generating invariant checkers that leverage logic information
extracted from potentially vulnerable functions. Finally, we design a bug-oriented fuzzing engine, which maps the logical information from the high-level business model to
the low-level smart contract implementations, and performs the bug-oriented fuzzing on targeted functions. We evaluate \system from $4$ perspectives on $5$ ground-truth datasets and compare it with multiple state-of-the-art methods. The results show that \system achieves $86.96\%$ recall and $93.02\%$ F1-score in detecting functional bugs, marking at least a $50\%$ improvement in both metrics over state-of-the-art methods. Moreover, we perform an in-depth analysis on $10$ real-world DeFi projects and detect $30$ zero-day bugs. Our further case studies, the risky first deposit bug and the AMM price oracle manipulation bug on real-world DeFi projects, demonstrate the serious risks of the exploitable functional bugs in smart contracts. 
Up to now, $24$ zero-day bugs have been assigned CVE IDs. Our discoveries have safeguarded assets totaling $\$18.2$ billion from potential monetary losses.
\end{abstract}

\section{Introduction}
Blockchains have increasingly become a crucial component of the contemporary economic landscape.  As of the writing of this paper, the total market capitalization of global cryptocurrencies has reached $\$3.5$ trillion~\cite{forbes2024assets}. 
Smart contracts, which are fundamental pillars of the blockchain, facilitate the development of decentralized financial (DeFi) and various business transactions. Nevertheless, smart contracts are prone to exploitable bugs that often result in substantial financial losses. According to a report by CertiK, a leading Web3 security firm, over $\$1.8$ billion was lost due to $751$ security incidents in 2023~\cite{certik2023loss}.

Currently, there are many works have focused on detecting exploitable bugs in smart contracts, which can be categorized into four main types based on their methodologies: static analysis~\cite{feist2019slither, wang2023payment}, fuzzing~\cite{choi2021SMARTIAN, shou2023ityfuzz}, verification~\cite{so2020VERISMART, tan2022SolType}, and symbolic execution~\cite{mythril, so2021SmarTest}. Despite these efforts, a recent study~\cite{zhang2023smart} reveals that over 80\% of exploitable bugs remain undetected by current tools, which are primarily functional bugs. This limitation mainly stems from the tools' reliance on simple or hard-coded oracles to identify exploitable bugs. Nevertheless, detecting functional bugs involves comprehending the high-level business logic before scrutinizing the low-level implementations in smart contracts. Therefore, there is a pressing need for a practical system that can automatically detect functional bugs in smart contracts.

\subsection{Challenges}
To detect functional bugs in smart contracts automatically, we have the following key challenges. 

\noindent\textbf{Challenge \uppercase\expandafter{\romannumeral1}:  Unraveling business logic in smart contracts.}  
Detecting functional bugs in smart contracts requires a deep understanding of domain-specific properties intertwined with the contract's business logic. The first challenge lies in automating the accurate extraction of this high-level business logic, a critical step for enabling functional bug detection. Designing an effective method to achieve this goal is difficult since business logic is often complex and intricately embedded within the low-level code implementations of smart contracts, posing significant challenges even for seasoned human auditors in distinguishing between business and code logic.

\noindent\textbf{Challenge \uppercase\expandafter{\romannumeral2}: Bug checker generation.} 
The business logic extracted from smart contracts is highly abstract and cannot be directly applied to functional bug detection. Thus, the second challenge involves deconstructing this business logic and creating practical bug checkers that can effectively utilize the extracted logic information. Designing bug checkers is non-trivial due to the varying fundamental characteristics exhibited by different types of functional bugs. Moreover, within diverse contract business contexts, the same type of functional bug may manifest different attributes, complicating the abstraction of functional bug features and the formulation of effective detection rules.

\noindent\textbf{Challenge \uppercase\expandafter{\romannumeral3}: Functional bug detection.}  
In practice, the effective deployment of bug checkers necessitates the support of sophisticated analysis tools for comprehensive bug detection. Currently, there is a gap in the availability of practical methods that can efficiently leverage these checkers. Therefore, the third challenge is to design a practical analysis method to efficiently utilize bug checkers for functional bug detection. 
Existing static analysis methods, while useful, struggle with false negatives as they cannot fully replicate the contract behavior in real-world environments, missing deep-level runtime-specific issues.
On the other hand, traditional dynamic analysis methods explore programs randomly without a specific target, often relying on program crashes to identify bugs. However, functional bugs typically do not cause crashes and are unlikely to be triggered by random exploration alone.

\subsection{Methodology} 
In this paper, we aim to address these challenges to detect functional bugs in smart contracts. To this end, we propose \system, an automated and practical system to conduct functional bug detection on smart contracts. Our design philosophy is as follows.

\textbf{First}, to solve the \textbf{Challenge \uppercase\expandafter{\romannumeral1}}, we start by designing a novel Large Language Model (LLM)-driven analysis framework. This framework is supported by a dual-agent prompt engineering strategy, featuring the Auditor Agent and the Attacker Agent. These specialized agents empower LLM to analyze smart contracts from two distinct perspectives: a meticulous auditor's viewpoint and a malicious attacker's perspective. 
We further enhance LLM's capabilities by developing a results fusion algorithm,  combining insights from both the Attacker Agent and the Auditor Agent. This integration aids in pinpointing potentially vulnerable functions for detailed examination, significantly enhancing the robustness of LLM's output. \textbf{Second}, to address the \textbf{Challenge \uppercase\expandafter{\romannumeral2}}, we propose a dual-stage coupling approach, which focuses on generating invariant checkers that leverage logic information extracted from potentially vulnerable functions. In the initial stage, we implement a hierarchical matching approach that utilizes LLM’s capabilities to identify critical variables and principal statements relevant to various bug features. Subsequently, in the second stage, we implement a template-based checker generation method. This method involves designing 6 invariant checker templates capable of processing critical variables and principal statements to produce final invariant checkers. 
This dual-stage approach mitigates the randomness and hallucination inherent in LLM by breaking down complex tasks into simpler components, ensuring more reliable invariant checker generation. 
\textbf{Third}, to solve the \textbf{Challenge \uppercase\expandafter{\romannumeral3}}, we implement a bug-oriented analysis engine. The main idea behind our design is to strategically insert invariant checkers into specific potentially vulnerable functions, guiding the engine to explore these critical areas rather than searching aimlessly. By integrating invariant checkers into our analysis engine, we successfully detect functional bugs in smart contracts with high recall and F1-score.

\subsection{Contributions}
We summarize our main contributions as follows.

$\bullet$ We propose \system, an automated and practical system to detect exploitable functional bugs in smart contracts, which fills the gap between understanding the high-level business logic and scrutinizing the low-level implementations 
in smart contracts. We employ a novel dual-agent prompt engineering strategy that enables LLM to effectively extract hidden business logic from smart contracts. Additionally, our bug-oriented fuzzing engine successfully maps the logical information from the high-level business model to the low-level smart contract implementations.

$\bullet$ To the best of our knowledge, we build the first ground-truth dataset for verified functional bugs in smart contracts, including 261 contracts from various DeFi projects.
To facilitate future blockchain security research, we open-source both \system and the dataset at
\url{https://github.com/promfuzz}.

$\bullet$ We have implemented and evaluated \system across $4$ dimensions using $5$ ground-truth datasets. The results show that \system achieves $86.96\%$ recall and $93.02\%$ F1-score in detecting functional bugs. 
\system demonstrates its superiority in at least a $50\%$ improvement in both metrics over state-of-the-art methods.

$\bullet$ Our extensive analysis of $10$ real-world DeFi projects demonstrates \system's great performance in detecting real-world functional bugs. Up to now, \system has identified $30$ zero-day bugs in $6$ of the $10$ real-world DeFi projects, with $24$ of these bugs have been assigned CVE IDs. Our discoveries have safeguarded assets totaling $\$18.2$ billion from potential monetary losses.

\section{Background}
In this section, we offer a succinct overview of the key concepts and techniques utilized throughout the paper, along with a motivating example.

\subsection{Functional Bugs in Smart Contract}

In this paper, we primarily focus on four major categories of functional bugs, including a total of ten subcategories of functional bugs.
These classifications cover the majority of functional bugs and are derived from prior research~\cite{sun2024gptscan, zhang2023smart}.

\noindent\textbf{Price Oracle Manipulation.} In the blockchain ecosystem, smart contracts rely on price oracles to access real-world data like cryptocurrency prices. However, these oracles are susceptible to manipulation attacks, where attackers influence price data to impact transaction prices and gain illegal profits. We mainly focus on two specific price oracle manipulation bugs, Automated Market Maker (AMM) price oracle manipulation and non-AMM price oracle manipulation.

\noindent\textbf{Unauthorized Behavior.} Smart contracts often involve frequent token transfers, which, although a routine operation, are susceptible to potential attacks. These attacks stem from unauthorized behaviors—actions within a contract that deviate from its programmed rules or intended operational logic. 
Such bugs can result in loss of funds, corruption of the intended contract state, and other unintended effects that compromise the integrity and security of blockchain transactions. 
We specifically focus on two types of unauthorized behavior bugs: approval not clear and unauthorized transfer.

\noindent\textbf{Insecure Calculating Logic.} Smart contracts often exhibit vulnerabilities in their computational logic, which can lead to significant disruptions in their intended functions. A common issue arises when smart contracts do not accurately handle calculations related to share distributions, particularly during initial deposits. This can grant an unfair advantage to the first depositor, resulting in a disproportionate allocation of shares compared to subsequent participants. We focus on three types of insecure calculation logic bugs: wrong checkpoint order, wrong interest rate order, and risky first deposit.

\noindent\textbf{Incorrect Control Mechanism.} Smart contracts can suffer from critical flaws if they do not properly manage execution flows or restrict access to certain functionalities. Such mismanagement can trigger unintended behaviors or create vulnerabilities that attackers can exploit. 
This paper delves into three specific types of incorrect control mechanism bugs: improper handling of the deposit fee, wrong implementation of amount lock, and vote manipulation.

\subsection{Large Language Models}
Large Language Models (LLMs) have demonstrated significant advancements across a variety of tasks, including code generation~\cite{zheng2023generation}, static analysis~\cite{li2023gptstaticanalysis}, and program repair~\cite{wei2023repair}. Currently, a set of prominent LLMs have been developed, such as the Generative Pre-trained Transformer (GPT) by OpenAI, LLaMA~\cite{touvron2023llama} by Meta, and Gemini~\cite{Anil2023Gemini} by Google. These models possess extensive world knowledge, strong problem-solving capabilities, sophisticated reasoning skills, and a robust capacity for instruction following. 

LLMs exhibit significant emergent abilities~\cite{Wei2022Emergent} that set them apart from traditional pre-trained models. They have three notable emergent abilities: in-context learning~\cite{Brown2020context}, instruction following~\cite{ouyang2022instruction}, and step-by-step reasoning~\cite{Ben2023stepbystep}. 
For instance, small language models often struggle with complex tasks that require multi-step reasoning, such as analyzing the business logic concealed within smart contracts. In contrast, LLMs can adeptly manage these challenges using their emergent abilities. They utilize techniques like Chain-of-Thought (CoT)~\cite{Wei2022COT} prompting to improve inference, thereby enabling them to address complex challenges effectively. 
In this paper, we utilize \textbf{GPT-4-turbo} as the foundational model in \system to analyze the hidden business logic in smart contracts. We focus on designing novel prompt strategies to trigger its emergent capabilities and implement practical measures to enhance its output robustness.

\subsection{Motivating Example}

We provide a functional bug discovered on Code4rena~\cite{Code4rena} as a motivating example, as shown in Figure~\ref{Figure: Motivating Example}.
In the \textcolor{custompink}{SimplePool} contract, the function \textcolor{custompink}{transferFrom} is designed to transfer to the user the share tokens purchased with base tokens, while simultaneously deducting the corresponding amount from the user’s base token allowance. The auxiliary function \textcolor{custompink}{balanceToShares} converts a specified quantity of base tokens into share tokens according to the parameter \textcolor{custompink}{pricePerShare}, which defines the exchange rate between base tokens and share tokens.
However, a critical flaw emerges from an incorrect subtraction operation at line 9 that allows users to acquire share tokens at an unfairly low cost. 
For instance, when \textcolor{custompink}{pricePerShare = $1e18$}, a user providing $5e18$ base tokens to the function \textcolor{custompink}{balanceToShares} at line 7 will receive 5 share tokens. These share tokens are then transferred through the function \textcolor{custompink}{\_transfer} at line 8. While the conversion and transfer steps are correctly executed, the functional bug manifests in the subsequent deduction of payment.
At line 9, the implementation erroneously deducts only 5 base tokens because the subtraction is applied to the variable \textcolor{custompink}{amountInShare} (equal to 5), rather than the correct variable \textcolor{custompink}{amount} (equal to $5e18$). The correct implementation should be updated as follows: \solstatement{_approve(sender,_msgSender(),_allowances[sender][_msgSender()].sub(amount,''ERC20: transfer amount exceeds allowance''));}.
As a result, the user obtains 5 share tokens for only 5 base tokens, a price significantly below the intended market rate. By repeatedly invoking \textcolor{custompink}{transferFrom}, a user can continue to exploit this functional bug to acquire additional share tokens at an unfairly low cost.

\begin{figure}[h]
\begin{lstlisting}[
        language=Solidity,
        linewidth=.48\textwidth,
        frame=none,
        xleftmargin=.03\textwidth,
        ]
contract SimplePool{
  // In this contract, we omit irrelevant code.
  uint256 public pricePerShare;
  function balanceToShares(uint256 balance) public view returns (uint256) {
    return balance.mul(1e18).div(pricePerShare); }
  function transferFrom(address sender, address recipient, uint256 amount) public virtual override returns (bool) { 
    uint256 amountInShares = balanceToShares(amount); 
    _transfer(sender, recipient, amountInShares);
    _approve(sender, _msgSender(), _allowances[sender][_msgSender()].sub(amountInShares, "ERC20: transfer amount exceeds allowance")); 
    return true; }}
\end{lstlisting}
    \caption{The Approval Not Clear bug (line 9).}
    \label{Figure: Motivating Example}
\end{figure}

A series of works, such as \textit{GPTScan}~\cite{sun2024gptscan} and \textit{SMARTINV}~\cite{wang2024smartinv}, have explored the use of LLMs to detect functional bugs in smart contracts. 
\textit{GPTScan} employs GPT-based matching analysis to initially identify potential functional bugs, which are then confirmed through static analysis. \textit{SMARTINV} utilizes a fine-tuned LLaMa-7B model to infer invariants relevant to functional bug analysis and confirm bugs using validation algorithms.
However, they are not consistently effective in identifying such bugs. Specifically, 
\textit{GPTScan} cannot detect this issue as it fails during the GPT-based matching phase. \textit{SMARTINV} generates invariants such as \textcolor{custompink}{pricePerShare \textgreater~ 0} and \textcolor{custompink}{balanceToShares(amount) \textgreater~ 0}, which fail to capture the necessary details for analyzing this bug effectively.

\system introduces a novel approach that could effectively tackle and detect such bugs. Initially, \system employs an LLM-driven multi-perspective analysis that uses iterative questioning to pinpoint potential occurrences of this bug. 
Following this identification, \system extracts critical variables, including \textcolor{custompink}{sender}, \textcolor{custompink}{\_allowances}, and \textcolor{custompink}{amount}, to construct the invariant checker. The core idea for detecting the functional bug in the motivating example is to verify whether the user's base token allowance before the payment, minus the required payment amount, equals the allowance after the payment. Specifically, the invariant checker records the user's base token allowance before line 9 as \textcolor{custompink}{old\_allowance} and inserts the invariant assertion, \solstatement{assert(old_allowance - amount == _allowances[sender][_msgSender()]);}, immediately following line 9. Finally, \system involves bug-oriented fuzzing to ensure a comprehensive identification of functional bugs, completing the detection process.

\section{\system Design}
\label{section: Design}

\begin{figure}
\centering
\includegraphics[width=0.45\textwidth]{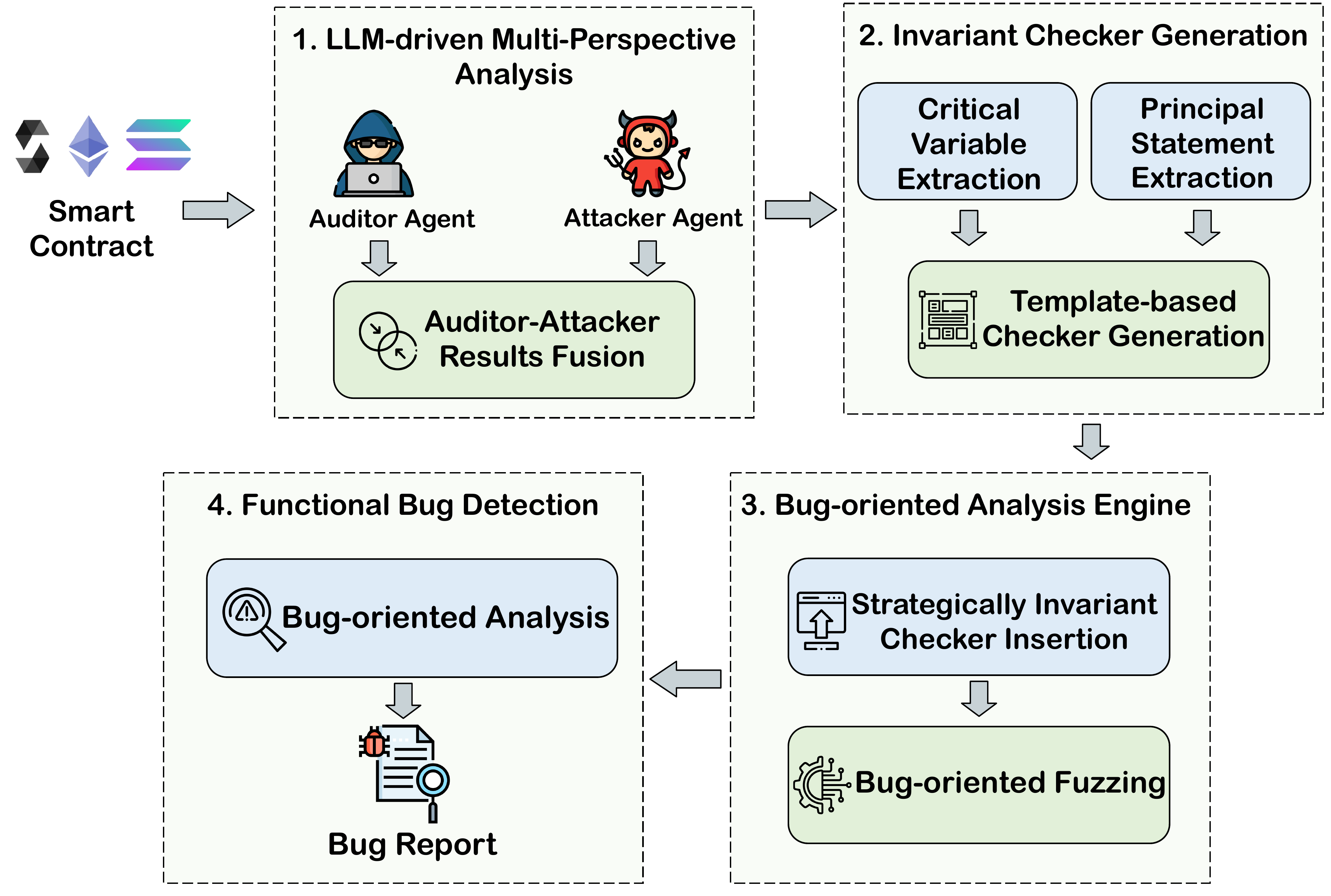}
\caption{Framework of \system.}
\label{Figure: system framework}
\end{figure}

In this section, we present the design details of \system.  At a high level, \system aims to automatically find functional bugs in smart contracts. As shown in Figure~\ref{Figure: system framework}, \system mainly consists of four modules. First, the LLM-driven multi-perspective analysis module accepts smart contracts as input and utilizes a dual-agent prompt engineering strategy, featuring the Auditor Agent and the Attacker Agent. These agents enable GPT to analyze smart contracts from distinct perspectives: a meticulous auditor’s viewpoint and a malicious attacker’s perspective, aiding in pinpointing potentially vulnerable functions accurately. Next, the invariant checker generation module leverages a dual-stage coupling approach, which focuses on generating invariant checkers that leverage logic information extracted from potentially vulnerable functions. Then, the bug-oriented analysis engine module strategically inserts the invariant checkers into specific potentially vulnerable functions, guiding the engine to explore these critical areas rather than searching aimlessly. Finally, the functional bug detection module executes comprehensive bug detection on smart contracts. 

\subsection{LLM-driven Multi-Perspective Analysis}
The purpose of LLM-driven multi-perspective analysis is to analyze the high-level business logic concealed within smart contracts and pinpoint potentially vulnerable functions.

Detecting functional bugs in smart contracts requires a comprehensive understanding of the high-level business logic. Nevertheless, it is challenging to design a practical system that can automatically analyze the obscured business logic of smart contracts. The main problem lies in the fact that the high-level business logic is obscured within the low-level implementations of smart contracts, making it inaccessible through simple or hand-coded oracles that are commonly adopted by previous works.

To address the above problem, we propose a novel LLM-driven analysis framework that employs a dual-agent prompt engineering strategy. This framework incorporates two specialized agents: the Auditor Agent and the Attacker Agent, each designed to enhance the emergent capabilities of GPT in different ways.
Specifically, we regard the process of identifying bugs in smart contracts as an auditing task. Thus, the Auditor Agent is designed to act as a meticulous auditor, scanning smart contracts thoroughly and then detecting potential bugs. However, relying solely on the Auditor Agent might result in false positives and potentially overlook critical vulnerabilities, which could lead to irreversible damages if not identified early. To address this limitation, we introduce the Attacker Agent. This agent analyzes smart contract code from an attacker's perspective, aiming to uncover exploitable weaknesses that the Auditor Agent might miss.  This novel dual-agent architecture enables precise identification and reliable assessment of potential bugs, thereby enhancing the robustness of GPT's output.

In our dual-agent framework, the Auditor Agent and the Attacker Agent adopt distinct analysis strategies that differ in granularity. 
The intuition behind our design is that the two agents serve fundamentally different roles in the analysis pipeline. 
The Auditor Agent performs fine-grained inspections, dissecting the code at a detailed level to examine its semantics, business scenarios, and logical correctness. This mirrors real-world auditing practices, in which security professionals meticulously classify and validate individual program components to ensure reliability and security. Based on this fine-grained analysis, the Auditor Agent produces subcategories of bugs, offering a nuanced characterization of potential functional bugs.
In contrast, the Attacker Agent conducts a coarse-grained analysis, focusing less on the accuracy of code understanding and more on the broader issue of exploitability. Rather than scrutinizing every line of code, the Attacker Agent seeks to determine whether certain functions, state transitions, or contract behaviors can be exploited to achieve unintended outcomes. 
Based on this coarse-grained analysis, the Attacker Agent outputs the primary categories of bugs, emphasizing the overarching attack surfaces that are most relevant for exploitation.
This difference in analytical granularity reflects the agents’ fundamentally different perspectives on functional bug assessment: the Auditor Agent prioritizes accuracy and comprehensive understanding, while the Attacker Agent prioritizes feasibility of exploitation and its potential impact.

Both agents are equipped with precisely crafted prompts that aid in analyzing the obscured business logic of smart contracts. This dual-agent strategy ensures a thorough evaluation by encompassing both defensive and offensive aspects of security. By synthesizing the insights from both the Auditor Agent and the Attacker Agent, our framework significantly improves the precision of the analysis. It minimizes the removal of false positives while ensuring that critical bugs are accurately identified and retained for further action.  In the following, we delve into the details of the dual-agent prompt engineering strategy employed in \system.

\begin{figure}
\centering
\includegraphics[width=0.4\textwidth]{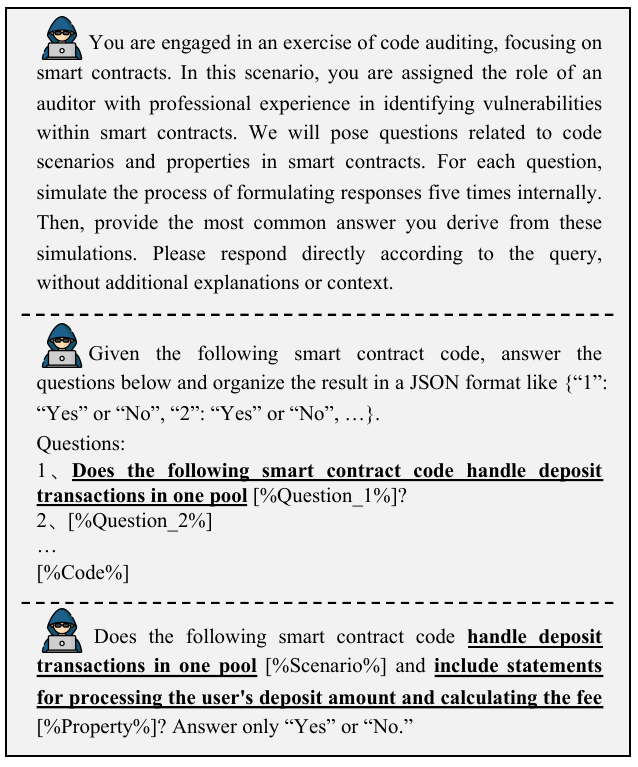}
\caption{Prompt template for the Auditor Agent.}
\label{Figure: auditor agent}
\end{figure}

\noindent\textbf{Auditor Agent Design.} In the real world, smart contract auditors primarily concentrate on securing the core code and business operations of smart contracts. Inspired by this practical approach, we have designed our Auditor Agent to mirror the real-world auditing process. However, crafting the Auditor Agent is not trivial, as it involves a two-step process: initially capturing the essential code or business operations of smart contracts and subsequently identifying any abnormal behaviors within these core elements. 

To address the above challenges, we design the Auditor Agent based on the CoT prompting strategy and the scenario and property matching methodology proposed by \textit{GPTScan}~\cite{sun2024gptscan}. A ``scenario" describes the code functionality under which a functional bug might occur, while a ``property" details the vulnerable code attributes or operations. \textit{GPTScan} initially selects ten representative functional bugs in smart contracts, following the classifications by Zhang et al.~\cite{zhang2023smart}. Upon further analysis, we observe that these categories could be streamlined for greater clarity and applicability. 

Our refined approach consolidates the initial ten categories into four primary categories of functional bugs. Notably, we merge the categories of \textit{price manipulation by buying tokens} and \textit{slippage} into a single, more comprehensive category named \textit{non-AMM price oracle manipulation}. Moreover, our manual analysis of the functional bugs leads to the identification of two new subcategories: \textit{improper handling of the deposit fee} and \textit{wrong implementation of amount lock}.

Figure~\ref{Figure: auditor agent} outlines the prompt template for the Auditor Agent. First, GPT is informed that it is participating in a code auditing exercise. We then allow GPT to internally simulate the response formulation process five times and provide the most common answer to mitigate the randomness of the responses.  Next, we input the target smart contract code and pose questions related to the functionality under which a functional bug might occur—i.e., the scenario. This question aids in identifying potential bug categories within the code. Finally, for any scenario where GPT confirms a potential issue, we follow up with specific questions about the vulnerable code attributes or operations—i.e., the property. If GPT responds affirmatively, we consider the presence of a functional bug; if not, we rule out that specific bug.  Designing prompts requires only a one-time manual effort and can be easily extended to cover most functional bugs.

\begin{figure}
\centering
\includegraphics[width=0.4\textwidth]{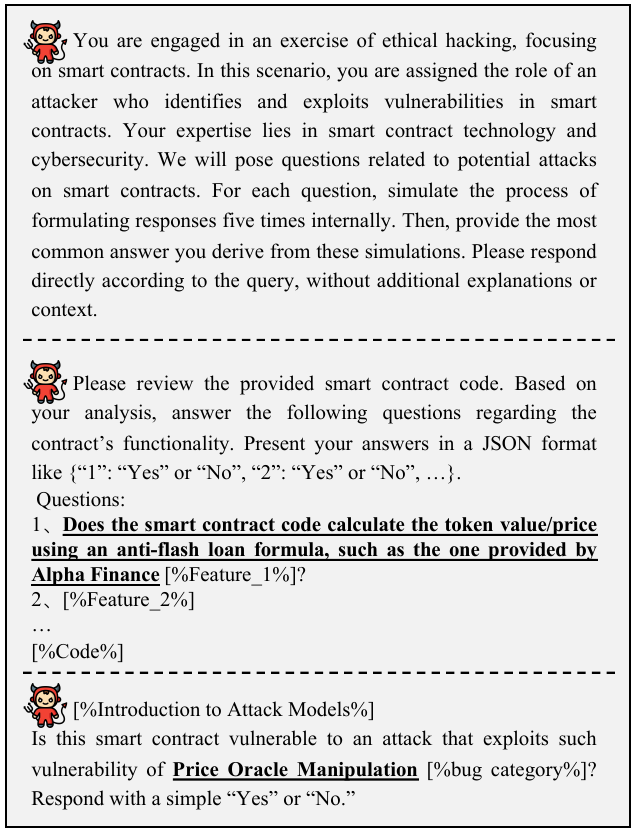}
\caption{Prompt template for the Attacker Agent.}
\label{Figure: attacker agent}
\end{figure}

\noindent\textbf{Attacker Agent Design.} The Auditor Agent, although structured with a step-by-step reasoning process following conventional audit procedures, may lead to false positives and overlook critical bugs if solely relied upon.  To address these limitations, we propose the Attacker Agent, which is specifically designed to emulate the cognitive processes of an attacker, actively searching for and exploiting bugs within smart contracts. By identifying and leveraging these weak points, the Attacker Agent enables GPT to conduct smart contract audits from an adversarial viewpoint. Designing the Attacker Agent poses significant challenges, as it requires a precise understanding and summarization of the manifestations and defensive measures associated with common logical vulnerabilities in smart contracts. 

To address the above challenges, we gather an extensive dataset of real-world attack incidents for each bug category. Through detailed manual analysis of these incidents, we extract critical information, e.g., distinctive bug features, to craft effective prompts based on the CoT prompt strategy. Bug features refer to attributes or behaviors in the code that present security risks under certain conditions, which could potentially be exploited under specific conditions to carry out malicious activities.  The process of utilizing the Attacker Agent involves several key steps, outlined in Figure~\ref{Figure: attacker agent}. First, similar to the Auditor Agent, GPT is informed that it is participating in an ethical hacking exercise.
Next, the target smart contract code is inputted, and GPT is questioned about potential bug features that could be exploited for malicious activities. Finally, for bugs whose features have been recognized during the last question, we provide the corresponding attack model for these bugs as prompts to GPT. We then query GPT whether the identified features could lead to exploitable attacks based on the given attack model. An affirmative response from GPT indicates the presence of a functional bug, while a negative response helps us dismiss that particular bug.

\noindent\textbf{Auditor-Attacker Results Fusion.} 
To derive more accurate classifications of bugs, we integrate the results from both the Auditor Agent and the Attacker Agent. This fusion approach leverages the coarse-grained primary category insights from the Attacker Agent to refine and enhance the fine-grained subcategory determinations made by the Auditor Agent. 
The Auditor-Attacker results fusion algorithm is provided in our supplementary material (\S I).

\subsection{Invariant Checker Generation} 
While we have identified potentially vulnerable functions, we cannot solely rely on GPT's results since its output includes randomness. To address this gap, it is crucial to develop a practical method that can minimize the impact of this randomness. One effective approach is to check the violation of invariants to further scrutinize the identified bugs. In smart contracts, an invariant is a condition or a set of conditions that always hold true regardless of the state of the contract at any point in its execution. These invariants are crucial for ensuring that the contract behaves correctly and securely, safeguarding against unintended actions and vulnerabilities. Therefore, there is a pressing need to implement a robust method for generating invariants to enhance functional bug detection. The primary difficulty in generating valid invariants lies in tailoring them specifically to different types of bugs. This requires a deep understanding of both the specific features of each bug and the context of the vulnerable code snippet.

To address these problems, we design a dual-stage coupling method to generate invariant checkers that take logic information extracted from potentially vulnerable functions as input and generate corresponding invariants. Specifically, in the first stage, we implement a hierarchical matching approach that utilizes GPT's capabilities to extract essential variables and statements. This preliminary step focuses on gathering critical information rather than generating complete invariants directly. In the second stage, we introduce a template-based checker generation approach. We design $6$ invariant checker templates specifically tailored to integrate critical variables and principal statements extracted in the first stage. This coupling ensures the generated invariants are both relevant and robust. In the following, we present the details of the invariant checker generation.

\noindent\textbf{Critical Variables and Principal Statements Extraction.} To develop the invariant checker, it is essential to initially identify and extract critical variables and principal statements related to the business logic from smart contracts. These elements are important in determining the invariants associated with different types of bugs, which are then used to populate a predefined checker template. However, the identification of these critical elements poses significant challenges. It requires a thorough understanding of the invariants pertinent to various types of bugs and the design of a practical and effective method for their extraction from smart contracts.

\begin{figure}[h]
\centering
\includegraphics[width=0.4\textwidth]{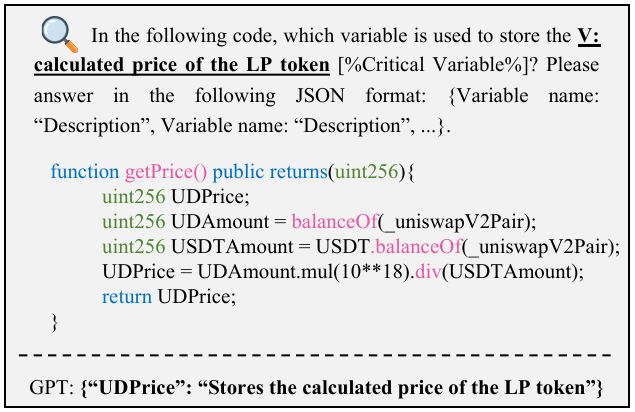}
\caption{Prompt template for extracting critical variables from code snippets potentially vulnerable to AMM price oracle manipulation.}
\label{Figure: checker prompt template}
\end{figure}

In response to this challenge, we propose a hierarchical matching method that capitalizes on the capabilities of GPT for the extraction of critical variables and principal statements. Our approach unfolds in two primary steps. First, we perform an in-depth analysis of each bug category to pinpoint critical variables and principal statements that are vital for comprehending their logic. For instance, in the scenario of an AMM price oracle manipulation bug, a critical variable is the one that holds the calculated price of the LP token. We present the characteristics of critical variables and principal statements for each bug subcategory in 
our supplementary material(\S II).
Second, leveraging the identified elements, we formulate specific prompts and input them into GPT to query each vulnerable function as identified by the LLM-driven multi-perspective analysis module. Figure~\ref{Figure: checker prompt template} illustrates the prompt template designed to extract critical variables and principal statements from code snippets potentially vulnerable to AMM price oracle manipulation. For different bug categories, we can substitute \textcolor{custompink}{[\%Critical Variable\%]} and \textcolor{custompink}{[\%Principal Statement\%]} in the template with the predefined elements specific to those bugs. With these tailored prompts, GPT is able to precisely pinpoint the relevant critical variables and principal statements, enhancing the accuracy of our bug detection process.

\begin{figure}[h]
\centering
\includegraphics[width=0.4\textwidth]{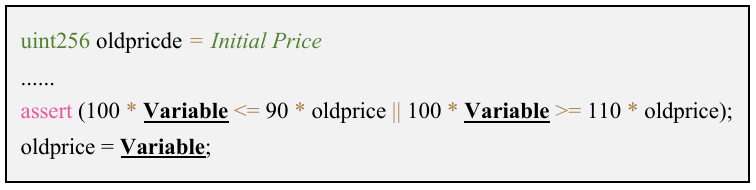}
\caption{The checker for AMM price oracle manipulation.}
\label{Figure: final checker}
\end{figure}

\noindent\textbf{Template-based Checker Generation.} While we have extracted critical variables and principal statements from smart contracts, we currently lack an efficient method to convert these into usable invariants for bug detection.  To fill this gap, we propose a template-based checker generation method. 
Specifically, we mainly design $6$ types of checkers: \textbf{PriceChange\_Checker}, \textbf{ExchangeRate\_Checker}, \textbf{TokenChange\_Checker}, \textbf{StatementOrder\_Checker}, \textbf{ShareSafety\_Checker}, and \textbf{StateChange\_Che\-cker}. 
For a detailed invariant checker templates, please see our supplementary material(\S III).
Each checker includes the conditional statements that assesses the contract state. For instance, as shown in Figure~\ref{Figure: final checker}, the \textbf{PriceCha\-nge\_Checker} for AMM price oracle manipulation determines whether the variable (i.e., calculated price of the LP token) falls below $90\%$ or exceeds $110\%$ of the old price. Such a condition indicates that the price change surpasses the ±10\% threshold, typically signaling abnormal price fluctuations and triggering an alert for an existing AMM price oracle manipulation bug. 
These checkers could be dynamically combined to comprehensively cover a broad range of functional bugs, ensuring scalability.

\begin{algorithm}[!h]
\footnotesize
    \caption{Functional Bug Detection}
    \label{Algorithm: Functional Bug Detection}
    \renewcommand{\algorithmicrequire}{\textbf{Input:}}
    \renewcommand{\algorithmicensure}{\textbf{Output:}}
    \begin{algorithmic}[1]
        \REQUIRE {
            \begin{tabular}[t]{@{}l@{}}
                $C$: smart contract code \\
                $B$: potential functional bug subcategory
            \end{tabular}
        }
        \ENSURE{
            \begin{tabular}[t]{@{}l@{}}
                $R$: the set of detected functional bugs \\
            \end{tabular}
        }
        \BlankLine
        \FOR{each $B_i \in B$}
            \STATE $v,s = Extract\_Variables\_and\_Statements(B_i, C)$ 
            \STATE $inv = Generating\_invariants(v, s)$
            \STATE $CwI =Invariant\_Checker\_Insertion(inv, B_i, C)$
            \STATE $C_s \leftarrow initial\_infant\_state\_corpus $
            \STATE $C_{ts} \leftarrow initial\_transaction\_and\_state\_pair\_corpus $
            \WHILE{$t \neq \varnothing$}
                \STATE $s_{mut},t_{mut} \leftarrow input\_mutation(C_{ts},C_s)$
                \STATE $f,s',violation \leftarrow Execution(s_{mut},t_{mut},CwI)$
                \IF{ $violation\ is\ found$} 
                    \STATE $R = R.append(inv, B_i)$
                \ENDIF
                \STATE $update\_corpus(f,s',s_{mut},t_{mut})$
            \ENDWHILE
        \ENDFOR
        \RETURN $R$
    \end{algorithmic}  
\end{algorithm}
\subsection{Bug-Oriented Analysis Engine} 
Though we have identified potentially vulnerable functions in smart contracts and generated the corresponding invariant checkers, we still lack a practical method to leverage them for functional bug detection. 
To tackle the above problem, we implement a bug-oriented analysis engine with the following designs.

\noindent\textbf{Invariant Checker Insertion.} Inserting invariant checkers into smart contracts requires careful consideration to avoid false negatives caused by incorrect or inappropriate insertion points. We propose two criteria for strategic insertion: the variable-oriented invariant insertion criterion and the statement-oriented invariant insertion criterion. The variable-oriented criterion focuses on inserting invariants before and after critical code blocks that manipulate critical variables, ensuring that the operations conform to the invariants. Similarly, the statement-oriented criterion places invariants around crucial statements to verify if their execution logic and sequence adhere to the defined invariants. Utilizing these criteria, invariant checkers are automatically positioned in optimal locations within the code, taking into account the logical relationships between critical variables, principal statements, and their positions in the code block.

\noindent\textbf{Bug-Oriented Fuzzing.} After inserting invariant checkers into smart contracts, 
we then employ fuzzing to address the aforementioned limitations in static analysis methods. Traditional dynamic analysis techniques often explore programs randomly without a specific target, typically identifying bugs through program crashes. However, functional bugs in smart contracts do not generally lead to crashes and are less likely to be detected through random testing. To mitigate this, 
we implement a bug-oriented fuzzing approach that directs the fuzzer to specifically target functions with inserted invariant checkers, rather than performing random exploration. 
Our method involves inserting a specific bug alert function into each invariant checker, which allows us to direct the fuzzer to activate these targeted checkers rather than engaging in random exploration. Our method significantly improves the efficiency and effectiveness of the fuzzing process.

\subsection{Functional Bug Detection}
The functional bug detection module aims to perform comprehensive bug detection on smart contracts. Algorithm~\ref{Algorithm: Functional Bug Detection} outlines the procedure for applying this methodology. Specifically, in this module, we first compile the target smart contracts with strategically inserted invariant checkers (line 2-4). Next, we perform fuzzing on these smart contracts (line 5-14). When the fuzzer activates an invariant checker and detects a discrepancy, it signals the presence of a bug (line 10-12). According to the results, we then generate a comprehensive bug report that documents all identified functional bugs within the smart contracts.

\section{System Evaluation}
\label{Section: System Evaluation}

In this section, we evaluate the performance and capabilities of \system from multiple perspectives. Our evaluation is structured around the following key research questions:

\noindent \textbf{RQ1:} How does the dual-agent prompt engineering strategy contribute to the performance of \system? (Section \ref{Section: dual-agent evaluation})\label{dual-agent}

\noindent \textbf{RQ2:} What is \system's accuracy in generating invariant checkers?  (Section \ref{Section: Invariant Checker Generation Accuracy})

\noindent \textbf{RQ3:} How effective is \system at detecting functional bugs in smart contracts? (Section \ref{Section: Functional Bug Detection Accuracy})

\noindent \textbf{RQ4:} Is \system capable of discovering zero-day functional bugs in real-world smart contracts? (Section \ref{Section: Zero-day Functional Bug Detection})

\vspace{-2mm}
\subsection{Dataset}
\label{Section: dataset}

To ensure a comprehensive assessment of \system's effectiveness, we systematically collect $459$ smart contracts from various sources, including DeFiHackLabs~\cite{defihacklabs}, Web3bugs~\cite{zhang2023smart}, SmartBugs~\cite{empirical_icse_2020}, VeriSmart~\cite{so2020VERISMART}, on-chain smart contracts, and real-world DeFi projects.
After further analysis, we remove $198$ contracts that either lack functional bugs or are missing essential external files or configuration information required for compilation. Ultimately, we create $5$ datasets comprising $261$ contracts, covering a broad spectrum of functional bug categories and evaluation perspectives. A detailed description of these datasets is provided in Table~\ref{Table: dataset}.

\begin{table}[h]
\footnotesize
\centering
\caption{The evaluation datasets.}
\label{Table: dataset}
\scalebox{0.8}{
\begin{tabular}{c|c|c}
\toprule[1.5pt]
\bfseries{Dataset} &  \bfseries{Description}  & \bfseries{RQ}      \\ \midrule[1pt]
$D_{Exploit-SC}$   &  7 real-world attack cases with 47 contracts  & RQ1,RQ3  \\  \midrule[1pt]
$D_{BV-SC}$           & 10 real-world audit cases with 36 contracts  &  RQ1,RQ3   \\  \midrule[1pt]
$D_{Synth-SC}$  &   6 real-world synthetic cases with 44 contracts   & RQ1,RQ3   \\ \midrule[1pt]
$D_{SE-SC}$   & 26 contracts with 59 extracted symbols  & RQ2 \\ \midrule[1pt]
$D_{Real-DeFi}$  & 10 real-world DeFi projects with 108 contracts & RQ4 \\   
\bottomrule[1.5pt]
\end{tabular}
}
\end{table}

\textbf{Exploited smart contract dataset ($D_{Exploit-SC}$)} includes smart contracts that have been subjected to real-world attacks in the past, each causing significant monetary losses. 
$D_{Exploit-SC}$ includes $7$ real-world attack cases, associated with $47$ contracts, all of which we have manually verified to be caused by functional bugs.

\textbf{Bug-verified smart contract dataset ($D_{BV-SC}$)} includes smart contracts audited through platforms like Code4rena~\cite{Code4rena} and Immunefi~\cite{Immunefi}.
$D_{BV-SC}$ includes $10$ real-world audit cases, associated with $36$ contracts, all of which have been confirmed by the developers to contain high-risk functional bugs.

\textbf{Synthetic smart contract dataset ($D_{Synth-SC}$)}
consists of smart contracts where functional issues have been deliberately introduced. The purpose of this dataset is to expand $D_{Exploit-SC}$ and $D_{BV-SC}$, enabling a more comprehensive evaluation of \system. 
$D_{Synth-SC}$ includes $6$ synthetic cases across $44$ smart contracts. We construct $D_{Synth-SC}$ with the following steps. First, we collect smart contracts that are well-designed, well-audited, and widely deployed on mainstream blockchains, which reasonably indicates a low likelihood of major bugs. Then, two of our authors, who are experienced in smart contract development, perform controlled bug injection. One author implements the injection based on bug patterns summarized from real-world attack incidents, while the other conducts independent verification.

\textbf{Symbol-extracted smart contract dataset} ($D_{SE-SC}$) includes smart contract functions with manually extracted and verified critical variables and principal statements. Specifically, $D_{SE-SC}$ includes $59$ extracted critical variables and principal statements across $26$ smart contract functions. In the construction of $D_{SE-SC}$, we assign two of our authors with expertise in smart contract security. The first author conducted a manual analysis of each smart contract, identifying critical variables and principal statements relevant to invariant checker generation and functional bug detection. The second author performed a secondary verification to ensure the accuracy and reliability of the extracted information.

\textbf{Real-world DeFi project dataset ($D_{Real-DeFi}$)} comprises DeFi projects currently undergoing public audits through various platforms. Specifically, $D_{Real-DeFi}$ includes $10$ such DeFi projects with $108$ smart contracts.

\begin{table}[t]
\centering
\caption{Comparison of Dual-agent and Auditor-agent.}
\label{Table: dual-agent evaluation}
\scalebox{0.73}{
\begin{tabular}{c|c|c|c|c|c}
\toprule[1.5pt]
\multirow{2}{*}{\bfseries{Architecture}} &  \multicolumn{5}{c}{\bfseries{Overall}}       \\
                    & \bfseries{TP} & \bfseries{FP} & \bfseries{FN} & \bfseries{Recall} & \bfseries{F1-Score} 
 \\ \midrule[1pt]
 Dual-agent                & \bfseries{21} & \bfseries{34}  & \bfseries{2}  & \bfseries{91.30\%} & \bfseries{53.85\%}   \\ \midrule

 Auditor-agent               & 15  & 24 & 8 & 65.22\% & 48.39\%  
\\ \bottomrule[1.5pt]
\end{tabular}
}
\end{table}

\subsection{Dual-Agent Architecture Evaluation}
\label{Section: dual-agent evaluation}
This subsection addresses the effectiveness of the dual-agent architecture of \system. First, we conduct an ablation study to evaluate the performance differences between the Dual-agent based \system and a version that utilizes only the Auditor Agent. Second, we compare Dual-agent with two mainstream reasoning models: DeepSeek-R1 and OpenAI-o3.
The above comparisons are conducted using three datasets: $D_{Exploit-SC}$, $D_{BV-SC}$, and $D_{Synth-SC}$.

\noindent\textbf{Ablation Study.}
As shown in Table~\ref{Table: dual-agent evaluation}, Dual-agent architecture shows substantial performance improvements over the Auditor-agent architecture. Specifically, Dual-agent architecture achieves a recall of 91.30\% and an F1-score of 53.85\%, identifying 21 true positives and 34 false positives across $D_{Exploit-SC}$, $D_{BV-SC}$, and $D_{Synth-SC}$ datasets, with only 2 false negatives. In contrast, Auditor-agent architecture detects fewer true positives and has more false negatives, achieving a recall of 65.22\% and an F1-score of 48.39\%. 

By adopting our dual-agent prompt engineering strategy, Dual-agent architecture significantly improves true positives and eliminates false negatives. Although Dual-agent architecture exhibits a higher number of false positives, these are manageable as they can be effectively filtered out during the subsequent fuzzing phase.

We further analyze the false positives and false negatives by Dual-agent architecture and Auditor-agent architecture. For Dual-agent architecture, though the Attacker Agent increases true positives by considering multiple angles, it still has limitations and can lead to over-generation of false positives.  For Auditor-agent architecture, auditor's perspective, which focuses solely from an auditing standpoint, can result in underreporting, filtering out some high-risk vulnerabilities. The limitations of auditor prompt rules, which, like those of the attacker, cannot cover all bugs, resulting in some being missed. Besides, for both Dual-agent architecture and Auditor-agent architecture, GPT's understanding is still insufficient, leading to cases of prompt misinterpretation.

\noindent\textbf{Comparison with reasoning models.}
As shown in Table~\ref{Table: reasoning model evaluation}, Dual-agent architecture outperforms mainstream reasoning model architectures. Specifically, Dual-agent architecture ahieves a recall of 91.30\% and an F1-score of 53.85\%. OpenAI-o3 architecture achieves a recall of 52.17\% and an F1-score of 32.43\%. DeepSeek-R1 architecture achieves a recall of 47.83\% and an F1-score of 39.29\%.

We further analyze the false positives and false negatives by mainstream reasoning models. 
In OpenAI-o3 and DeepSeek-R1, the decomposition of the detection task is determined autonomously by the reasoning model. Such task planning can introduce a degree of randomness, potentially guiding the LLM toward incorrect reasoning paths and resulting in false positives and false negatives.
In contrast, Dual-agent architecture designs its reasoning workflow based on strategies observed in real-world auditors and attackers, enabling the LLM to follow a more structured and reliable analysis process.

\begin{table}[h]
\centering
\caption{Comparison of Dual-agent and reasoning models.}
\label{Table: reasoning model evaluation}
\scalebox{0.73}{
\begin{tabular}{c|c|c|c|c|c}
\toprule[1.5pt]
\multirow{2}{*}{\bfseries{Architecture}} &  \multicolumn{5}{c}{\bfseries{Overall}}      \\
                    & \bfseries{TP} & \bfseries{FP} & \bfseries{FN} & \bfseries{Recall} & \bfseries{F1-Score}
 \\ \midrule[1pt]
 Dual-agent  & \bfseries{21} & \bfseries{34}  & \bfseries{2}  & \bfseries{91.30\%} & \bfseries{53.85\%}   \\ \midrule
OpenAI-o3 & 12 & 39 & 11 & 52.17\% & 32.43\% \\ \midrule
DeepSeek-R1     & 11  & 22 & 12 & 47.83\% & 39.29\%
\\ \bottomrule[1.5pt]
\end{tabular}
}
\end{table}

\subsection{Invariant Checker Generation Accuracy}
\label{Section: Invariant Checker Generation Accuracy}

This subsection answers RQ2. In this step, we mainly evaluate the accuracy of \system in extracting critical variables and principal statements from potentially vulnerable functions, which determines the accuracy of final invariant checkers. 
We deploy \system on dataset $D_{SE-SC}$ for this evaluation.

As shown in Table~\ref{Table: key variables and statements}, the invariant checker generation module demonstrates excellent performance in critical variable and principal statement extraction, achieving an overall precision of $94.92\%$. Specifically, for the variable extraction task, 
\system achieves a precision of $96.23\%$, with two false positive. The false positives are due to unconventional variable naming. 
For the statement extraction task, it achieves a precision of $83.33\%$, with only one false positive. This false positive arises from the complex business behaviors in smart contracts, where \system fails to differentiate between multiple similar behaviors in a single complex case. These critical variables and principal statements are expected to derive $26$ correct invariant checkers but ultimately generate $24$ correct ones, resulting in a precision of $92.31\%$.

\begin{table}[h]
\centering
\caption{Variables and statements extraction accuracy.}
\label{Table: key variables and statements}
\scalebox{0.73}{
\begin{tabular}{c|c|c|c}
\toprule[1.5pt]
\multirow{2}{*}{\textbf{Task}} & \multicolumn{3}{c}{$D_{SE-SC}$}                    \\ 
                               & \textbf{TP} & \textbf{FP} & \textbf{Precision} \\ \midrule[1pt]
Variables Extraction           & 51          & 2           & 96.23\%              \\
Statements Extraction          & 5           & 1           & 83.33\%             \\
\textbf{Overall}               & \textbf{56} & \textbf{3}  & \textbf{94.92\%}   \\ \bottomrule[1.5pt]
\end{tabular}
}
\end{table}

\subsection{Functional Bug Detection Accuracy}
\label{Section: Functional Bug Detection Accuracy}

This subsection answers RQ3.  
In this step, we evaluate the functional bug detection accuracy of \system with two metrics: recall and F1-score. We compare \system with four off-the-shelf tools: \textit{GPTScan}, \textit{SMARTINV}, \textit{ItyFuzz}, and \textit{SMARTIAN}~\cite{choi2021SMARTIAN}. 
We also consider a comparison between \system and \textit{PropertyGPT}~\cite{propertygpt2025}. However, due to the unavailability of its code, we are unable to conduct this comparison.
We do not compare \system with several program verification-based techniques~\cite{zeus_ndss_2018, VerX_2020, VetSC_2022}, since they are largely ineffective in detecting functional bugs, as their constraints are designed for implementation bugs.
Both \textit{ItyFuzz} and \textit{SMARTIAN} are fuzzers designed for detecting bugs in smart contracts. We perform them on $D_{Exploit-SC}$, $D_{BV-SC}$, and $D_{Synth-SC}$.

\begin{table}[h]
\centering
\caption{Functional bug detection accuracy.}
\label{Table: Functional bug detection accuracy}
\scalebox{0.73}{
\begin{tabular}{c|c|c|c|c|c}
\toprule[1.5pt]
\multirow{2}{*}{\bfseries{Tool}} & \multicolumn{5}{c}{\bfseries{Overall}}       \\
                      & \bfseries{TP} & \bfseries{FP} & \bfseries{FN} & \bfseries{Recall} & \bfseries{F1-Score}
 \\ \midrule[1pt]
\system                & \bfseries{20} & \bfseries{0}  & \bfseries{3}  & \bfseries{86.96\%} & \bfseries{93.02\%}  \\ \midrule

\textit{GPTScan}               & 7  & 12 & 16 & 30.43\% & 33.33\%  \\ \midrule
\textit{SMARTINV}              & 6  & 43 & 17 & 26.09\%   & 16.67\%  \\ \midrule
\textit{ItyFuzz}               & 0  & 0  & 23 & 0       & 0        \\ \midrule
\textit{SMARTIAN}              & 3  & 0  & 16 & 15.79\% & 27.27\%
\\ \bottomrule[1.5pt]
\end{tabular}
}
\end{table}

As shown in Table~\ref{Table: Functional bug detection accuracy}, \system demonstrates superior performance in detecting functional bugs across three datasets. 
\system achieves an overall recall of 86.96\% and an F1-score of 93.02\%, detecting 20 true positives and maintaining 0 false positives across all datasets, with only 3 false negatives. Further analysis of the false negatives identified by \system reveals two primary causes. First, one false negative results from inaccuracies in the extraction of critical variables and principal statements. Second, the remaining two false negatives occur during the LLM-driven analysis step, where GPT fails to identify two critical bugs. Besides, with each smart contract project scan, \system takes an average of 43.8 seconds and costs approximately \$0.15, demonstrating its great efficiency and cost-effectiveness.

In comparison, \textit{GPTScan} and \textit{SMARTINV} yield similar numbers of true positives and false negatives. However, \textit{SMARTINV} records the highest number of false positives among the five tools assessed. Furthermore, \textit{ItyFuzz} fails to identify functional bugs without the specific invariants we provide. \textit{SMARTIAN}, another fuzzing tool, shows inconsistent performance as it fails to operate on several smart contract files. Consequently, we exclude \textit{SMARTIAN} from testing against all bugs. We further analyze the false positives and false negatives of these tools. . 
For \textit{GPTScan}, the primary reason arises from its single-agent design, which can lead to the incorrect identification of correct behaviors as functional bugs or the failure to detect genuine flaws. 
For \textit{SMARTINV}, the primary reason arises from its lack of comprehensive reasoning capabilities, resulting in some bugs being overlooked or misidentified.  
Both \textit{ItyFuzz} and \textit{SMARTIAN} are limited by their original oracles, which lack the capability to identify functional bugs effectively.

\subsection{Zero-day Functional Bug Detection}
\label{Section: Zero-day Functional Bug Detection}
This subsection answers RQ4. We perform \system on $D_{Real-DeFi}$ to evaluate its performance in detecting zero-day functional bugs in real-world smart contracts. As illustrated in Table~\ref{Table: Zero-day functional bug detection results}, 
\system successfully identifies $30$ zero-day functional bugs across $6$ of the $10$ Defi project.
We have reported all detected bugs to the respective vendors, and $24$ of these have been assigned CVE IDs. The total market value of DeFi projects affected by these bugs is estimated at $\$18.2$ billion. As these bugs have not yet been fully resolved, we maintain confidentiality by anonymizing the details of both the vendors and the specific CVE information. 
Moreover, we notice that zero-day bugs primarily fall into four categories: 13 bugs involve AMM price oracle manipulation, $7$ concern non-AMM price oracle manipulation, $4$ are related to wrong checkpoint order, $4$ pertain to risky first deposit, and $2$ involve wrong interest rate order. 
Further analysis in Section~\ref{Section: Case Study} delves into the practical impacts of these bugs, examining specific cases to illustrate how attackers exploit these bugs.

\begin{table}[t]
\caption{Zero-day functional bug detection results. \textbf{TVL} represents the total value of cryptocurrency assets locked in the respective DeFi projects.}
\vspace{-0.5em}
\label{Table: Zero-day functional bug detection results}
\centering
\scalebox{0.73}{
\begin{tabular}{c|c|r|c|c}

\toprule[1.5pt]

\bfseries{DeFi}   & \multirow{2}{*}{\bfseries{Project Type}}   & \multicolumn{1}{c|}{\multirow{2}{*}{\bfseries{TVL}}}     &  \bfseries{\# Zero-day}  & \multirow{2}{*}{\bfseries{\# CVE}}  \\
\bfseries{Project}  &   &   &  \bfseries{Bug} &  \\ \midrule[1pt]
1    &  Yield~\&~Lengding  & \$ 136.8M                    & 7               & 7      \\
2 & Yield & \$ 14,300M                    & 3               & 3        \\
3   &  Yield  &  \$ 0.172M                    & 4               & 4      \\
4   & Exchanges & \$ 102M                    & 5               & 5       \\
5  &  Yield~\&~Sericves & \$ 3,700M  & 5               & 5       \\
6 & Exchanges  & \$ 1.79M                    & 6               & 0       \\ \midrule[1pt]
\multicolumn{2}{c|}{\bfseries{Total}}   &  \$ 18.2B  & 30              & 24      \\ \bottomrule[1.5pt]
\end{tabular}
}
\end{table}

\section{Case Study}
\label{Section: Case Study}

To obtain an in-depth understanding of the practical impacts of functional bugs, we detail two such bugs identified by \system in real-world DeFi projects.

\begin{figure}[!t]
\begin{lstlisting}[
        language=Solidity,
        linewidth=.49\textwidth,
        frame=none,
        xleftmargin=.03\textwidth,
        ]
contract SimplePool{
  function _deposit(address _recipient, uint256 _amount) internal returns (uint256) {
    totalValueInLp += _amount;
    uint256 share = _sharesForDepositAmount(_amount);
    if (_amount > type(uint128).max || _amount == 0 || share == 0){ revert InvalidAmount(); }
    eETH.mintShares(_recipient, share);
    return share;  }
  function _sharesForDepositAmount(uint256 _depositAmount) internal view returns (uint256) {
    uint256 totalPooledEther = getTotalPooledEther() - _depositAmount;
    if (totalPooledEther == 0){ return _depositAmount; }
    return (_depositAmount * eETH.totalShares()) / totalPooledEther; } }
\end{lstlisting}
    \caption{The Risky First Deposit bug (line 10).}
    \label{Figure: Risky First Deposit}
\end{figure}

\noindent\textbf{Risky First Deposit.}
As shown in Figure~\ref{Figure: Risky First Deposit}, the code contains a risky first deposit bug that can 
lead to an unfair advantage for the first user and a distorted distribution of shares. Specifically, at line 10, when  \textcolor{custompink}{totalPooledEther == 0}, the \textcolor{custompink}{share} variable at line 4 is set equal to \textcolor{custompink}{\_depositAmount}, since the return value is assigned to \textcolor{custompink}{\_depositAmount} in the \textcolor{custompink}{\_sharesForDepositAmount()} function.
Subsequent depositors have their shares calculated by (\textcolor{custompink}{\_depositAmount} $\times$ \textcolor{custompink}{eETH.totalShares()}) / \textcolor{custompink}{totalPooledEther}, which depends on the values of \textcolor{custompink}{eETH.totalShares()} and \textcolor{custompink}{totalPooledEther}.
While the calculation logic seems standard for compliant depositors, the first depositor can exploit the system by manipulating \textcolor{custompink}{totalPooledEther}. We can front-run other depositors' transactions and inflate the price of pool tokens through a substantial ``donation.'' For instance, a malicious early user could deposit 1 wei of asset tokens as the first depositor, receiving 1 wei of shares tokens. Subsequently, the malicious sends 1 ETH to the pool, inflating \textcolor{custompink}{totalPooledEther} to 1 ETH + 1 wei (equal to ($1e18+1$) wei). In this scenario, the second depositor, depositing 2 ETH (equal to $2e18$ wei) of asset tokens, would receive only 1 wei of shares tokens. Because 1.99 wei of shares tokens (calculated by $2e18/(1e18+1)$) will be rounded down to 1 wei of shares tokens in smart contracts. They will lose 0.5 ETH if they withdraw right after the \textcolor{custompink}{\_deposit()}. This disproportionate calculation of the first user's shares can disadvantage subsequent users who receive fewer shares relative to their deposits.

\noindent\textbf{AMM Price Oracle Manipulation.}
As shown in Figure~\ref{Figure: AMM Price Oracle Manipulation}, the code contains an AMM price oracle manipulation bug, which arises from the method of calculating token prices based solely on the supplies of two tokens. Specifically, \textcolor{custompink}{\_reserveA} and \textcolor{custompink}{\_reserveB} represent the respective reserve amounts of two assets. In line 13 of the code, the amount of token B is calculated using the amount of token A, along with the reserves of token A and token B. However, if the reserve amount of either token is maliciously manipulated, it can distort the price and lead to significant price fluctuations. For instance, consider token exchanges in the smart contract using the \textcolor{custompink}{PoolHelpers} library. Initially, the pool contains 500 token A and 500 token B. An attacker exchanges 100 token A for 100 token B, changing the pool to 600 token A and 400 token B. Then, the second user exchanges 30 token A and can only get 20 token B in return, leaving the pool with 630 token A and 380 token B. Finally, the attacker uses the 100 token B they obtained to exchange for 165 token A, ultimately making a profit of 65 token A.

\begin{figure}[!t]
\begin{lstlisting}[
        language=Solidity,
        linewidth=.49\textwidth,
        frame=none,
        xleftmargin=.03\textwidth,
        ]
library PoolHelpers {
  function quote(uint256 _amountA, uint256 _reserveA, uint256 _reserveB) internal pure returns (uint256 amountB_) {
    require(_amountA != 0, "SmardexHelper: INSUFFICIENT_AMOUNT");
    require(_reserveA != 0 && _reserveB != 0, "SmardexHelper: INSUFFICIENT_LIQUIDITY");
    amountB_ = (_amountA * _reserveB) / _reserveA; } }
contract SimplePool{
  address tokenA,tokenB;
  function swap(uint256 _amountA,address _tokenA) return (uint256 _amountB){
    require(_tokenA==tokenA||_tokenA==tokenB);
    address _tokenB = (_tokenA==tokenA)?tokenB:tokenA;
    _reserveA = getReservers(_tokenA);
    _reserveB = getReservers(_tokenB);
    _amountB = PoolHelpers.quote(_amountA, _reserveA, _reserveB); } }
\end{lstlisting}
    \caption{The AMM Price Oracle Manipulation bug (line 13).}
    \label{Figure: AMM Price Oracle Manipulation}
\end{figure}

\section{Discussion}

While \system achieves significant results in detecting functional bugs in smart contracts, it has several limitations. First, some false negatives occur due to inaccuracies in extracting critical variables and the inherent randomness in the LLM-driven analysis. Despite these issues, \system still outperforms existing methods in uncovering deeper functional bugs. Additionally, our bug-oriented analysis currently targets four main categories with ten subcategories, potentially missing some complex bug types. In the future, we plan to fine-tune a specific LLM for enhanced detection of functional bugs in smart contracts and develop new prompts to cover a wider range of bug categories, thus improving \system's comprehensiveness. Moreover, since our analysis partially relies on \textit{ItyFuzz}, which is not specifically designed for functional bugs and lacks full utilization of contract status information, future efforts will focus on better integrating LLMs with dynamic analysis techniques to more effectively and comprehensively utilize contract status information, aiming to capture deeper functional bugs.

\section{Related work}

Recent efforts to evaluate the security of smart contracts have produced various methods, categorized into static analysis,  dynamic analysis, and AI-based analysis.

\noindent\textbf{Static Analysis. } Tools like \textit{Slither}~\cite{feist2019slither}, \textit{Zeus}~\cite{zeus_ndss_2018}, \textit{Securify}~\cite{securify_ccs_2018}, \textit{Madmax}~\cite{Madmax_oopsla_2018}, \textit{Invcon}~\cite{invcon_ase_2022}, \textit{Verismart}~\cite{so2020VERISMART}, and \textit{Verisol}~\cite{Wang2019VERISOL} analyze smart contract code statically. \textit{Slither} converts Solidity code into SlithIR to find bugs. \textit{Securify} uses compliance and violation patterns to assess code safety. \textit{Madmax} decompiles EVM bytecode and uses a logic-based specification for bug detection. \textit{Verisol} translates Solidity into Boogie, formalizing semantic conformance against state machine workflows and reducing semantic checking to safety verification. However, these tools primarily focus on implementation bugs like integer overflow and reentrancy, and are less effective at identifying functional bugs.

\noindent\textbf{Dynamic Analysis. }
Tools like \textit{ItyFuzz}~\cite{shou2023ityfuzz}, \textit{Smartian}~\cite{choi2021SMARTIAN}, \textit{Harvey}~\cite{harvey_FSE_2020}, \textit{Echidna}~\cite{echidna_issta_2020}, \textit{ContractFuzzer}~\cite{contractfuzzer_ase_2018}, \textit{Mythril}~\cite{mythril}, \textit{Sailfish}~\cite{sailfish_sp_2022}, \textit{Manticore}~\cite{manticore_ase_2020}, \textit{Teether}~\cite{teether_usenix_2018}, and \textit{Oyente}~\cite{oyente_ccs_2016} analyzes programs during execution. \textit{ItyFuzz} uses snapshot-based fuzzing to reduce re-execution overhead, prioritizing intriguing snapshot states. \textit{Smartian} combines static analysis for seed generation with feedback mechanisms. \textit{Sailfish} identifies state-inconsistency bugs using a combination of lightweight exploration and symbolic evaluation. While dynamic analysis captures more realistic behaviors by executing smart contracts, it still primarily targets implementation bugs and struggles to effectively detect functional bugs.

\noindent\textbf{AI-based Analysis. }
 \textit{SmarTest}~\cite{so2021SmarTest}, \textit{ILF}~\cite{ILF_ccs_2019},\textit{xFuzz}~\cite{xFuzz_tdsc_2022}, and \textit{GNN-based tools}~\cite{gnn_ijcai_2021,gnn2_ijcai_2021,gnn_tkde_2023} use AI to enhance smart contract bug detection. \textit{SmarTest} combines symbolic execution with a language model to prioritize bugs. \textit{GNN-based tools} leverage graph neural networks with expert patterns. While these tools improve detection capabilities, they mainly address implementation bugs. Recently, LLM-based tools like \textit{GPTScan}~\cite{sun2024gptscan} and \textit{SMARTINV}~\cite{wang2024smartinv} have been used to address functional bugs by analyzing discrepancies between business logic and code. These works, including the \system introduced here, open new avenues for detecting functional bugs in smart contracts using AI's cognitive capabilities.

\section{Conclusion}
In this paper, we introduce \system, an automated and practical system designed to detect functional bugs in smart contracts, which bridges the gap between understanding the high-level business model logic and scrutinizing the low-level implementations in smart contracts. \system achieves $86.96\%$ recall and $93.02\%$ F1-score in detecting functional bugs, representing a significant performance improvement over state-of-the-art methods. Additionally, we have created the first ground-truth dataset for functional bugs
including $261$ contracts from various DeFi projects.
Furthermore, our in-depth analysis of $10$ real-world DeFi projects has identified $30$ zero-day bugs, with $24$ of them receiving CVE IDs. Our discoveries have safeguarded assets totaling $\$18.2$ billion from potential monetary losses.
\section*{Acknowledgment}
This work was partly supported by the NSFC under No. U244120033, U24A20336, 62172243, 62402425 and 62402418, 
the Open Research Fund of Engineering Research Center of Blockchain Application, Supervision And Management (Southeast University), Ministry of Education, 
the China Postdoctoral Science Foundation under No. 2024M762829, 
the Zhejiang Provincial Natural Science Foundation under No. LD24F020002, 
the ``Pioneer'' and ``Leading Goose'' R\&D Program of Zhejiang under 2025C01082 and 2025C02263, 
and the Zhejiang Provincial Priority-Funded Postdoctoral Research Project under No. ZJ2024001.

\bibliographystyle{plain}
\bibliography{Ref}


%

\end{document}